\documentclass[modern]{aastex631}

\usepackage{xspace}

\newcommand{\nicer}{\textit{NICER}\xspace}
\newcommand{\pint}{\textsc{Pint}\xspace}
\newcommand{\enterprise}{\textsc{Enterprise}\xspace}
\newcommand{\ptmcmc}{\textsc{PTMCMCSampler}\xspace}
\newcommand{\tempotwo}{\textsc{Tempo2}\xspace}
\newcommand{\libstempo}{\textsc{libstempo}\xspace}
\newcommand{\hasasia}{\texttt{hasasia}\xspace}
\newcommand{\ftwo}{$\ddot{f}$\xspace}

\newcommand{\psrj}{\text{PSR J1824$-$2452A}\xspace}
\newcommand{\psrb}{\text{PSR B1937+21}\xspace}

\shorttitle{NICER Red Noise}
\shortauthors{Hazboun et al.}
\graphicspath{{./}{images/}}

\begin{document}

\title[NICER Red Noise Modeling]{A Detection of Red Noise in \psrj and Projections for \psrb using NICER X-ray Timing Data}

\correspondingauthor{Jeffrey S. Hazboun}
\email{hazboun@uw.edu}

\author[0000-0003-2742-3321]{Jeffrey S. Hazboun}
\affiliation{University of Washington Bothell, 18115 Campus Way NE, Bothell, WA 98011}

\author{Jack Crump}
\affiliation{Haverford College, 370 Lancaster Ave, Haverford, PA 19041, USA}

\author[0000-0003-4137-7536]{Andrea N. Lommen}
\affiliation{Haverford College, 370 Lancaster Ave, Haverford, PA 19041, USA}

\author{Sergio Montano}
\affiliation{Haverford College, 370 Lancaster Ave, Haverford, PA 19041, USA}

\author{Samantha J. H. Berry}
\affiliation{Bryn Mawr College, 101 North Merion Ave, Bryn Mawr, PA 19010, USA}

\author{Jesse Zeldes}
\affiliation{Haverford College, 370 Lancaster Ave, Haverford, PA 19041, USA}

\author{Elizabeth Teng}
\affiliation{Center for Interdisciplinary Exploration and Research in Astrophysics (CIERA), Northwestern University, Evanston, IL 60208, USA}
\affiliation{Haverford College, 370 Lancaster Ave, Haverford, PA 19041, USA}

\author[0000-0002-5297-5278]{Paul S. Ray}
\affiliation{U.S. Naval Research Laboratory, Washington, DC 20375, USA}

\author[0000-0002-0893-4073]{Matthew Kerr}
\affiliation{U.S. Naval Research Laboratory, Washington, DC 20375, USA}

\author{Zaven Arzoumanian}
\affiliation{Astrophysics Science Division, 
NASA Goddard Space Flight Center, Greenbelt, MD 20771, USA}

\author[0000-0002-9870-2742]{Slavko Bogdanov}
\affiliation{Columbia Astrophysics Laboratory, Columbia University,
550 West 120th Street, New York, NY 10027, USA}

\author{Julia Deneva}
\affiliation{George Mason University,
resident at Naval Research Laboratory, Washington, DC 20375, USA}

\author[0000-0003-0771-6581]{Natalia Lewandowska}
\affiliation{Department of Physics and Astronomy,
Swarthmore College, Swarthmore, PA 19081, USA}

\author{Craig B. Markwardt}
\affiliation{Astrophysics Science Division, 
NASA Goddard Space Flight Center, Greenbelt, MD 20771, USA}

\author[0000-0001-5799-9714]{Scott Ransom}
\affiliation{NRAO, 520 Edgemont Rd., Charlottesville, VA 22903, USA}

\author[0000-0003-1244-3100]{Teruaki Enoto}
\affiliation{Department of Astronomy, Kyoto University,
Kitashirakawa-Oiwake-cho, Sakyo-ku, Kyoto 606-8502, Japan}

\author{Kent S. Wood}
\affiliation{Praxis, resident at the Naval Research Laboratory, Washington, DC 20375, USA}

\author{Keith C. Gendreau}
\affiliation{Astrophysics Science Division, 
NASA Goddard Space Flight Center, Greenbelt, MD 20771, USA}

\author{David A. Howe}
\affiliation{Time and Frequency Division, NIST Boulder, CO 80305, USA}
\affiliation{Department of Physics, University of Colorado Boulder, CO 80309, USA}

\author[0000-0002-4140-5616]{Aditya Parthasarathy}
\affiliation{Max-Planck-Institut f\"ur Radioastronomie, Auf dem H\"ugel 69, D-53121 Bonn, Germany}


\begin{abstract}
We have used X-ray data from the Neutron Star Interior Composition Explorer (\nicer) to search for long time-scale, correlated variations (``red noise'') in the pulse times of arrival from the millisecond pulsars \psrj and \psrb. These data more closely track intrinsic noise because X-rays are unaffected by the radio-frequency dependent propagation effects of the interstellar medium. Our Bayesian search methodology yields strong evidence (natural log Bayes factor of $9.634 \pm 0.016$) for red noise in \psrj, but is inconclusive for \psrb. In the interest of future X-ray missions, we devise and implement a method to simulate longer and higher precision X-ray datasets to determine the timing baseline necessary to detect red noise. We find that the red noise in \psrb can be reliably detected in a 5-year mission with a time-of-arrival (TOA) error of 2 microseconds and an observing cadence of 20 observations per month compared to the 5 microsecond TOA error and 11 observations per month that \nicer currently achieves in \psrb. We investigate detecting red noise in \psrb with other combinations of observing cadences and TOA errors. We also find that an injected stochastic gravitational wave background (GWB) with an amplitude of $A_{\rm GWB}=2\times10^{-15}$ and spectral index of $\gamma_{\rm GWB}=13/3$ can be detected in a pulsar with similar TOA precision to \psrb, but with no additional red noise, in a 10-year mission that observes the pulsar 15 times per month and has an average TOA error of 1 microsecond.

\end{abstract}


\section{Introduction}
Rotation-powered pulsars, particularly millisecond pulsars (MSPs), are extraordinarily stable rotators, but there is noise in the arrival time of the pulses. Pulsar timing noise is the subject of much research, both to understand the pulsar emission mechanism, and for the sake of using pulsars as clocks, e.g. to make a gravitational wave detector, i.e. a pulsar timing array (PTA) \citep{haa+10,Arzoumanian_2016,Shannon_2012,Lentati:2016ygu,IPTADR2}.  The noise we observe likely comes from both intrinsic and extrinsic effects, but what fraction of the noise is intrinsic to the pulsar is difficult to disentangle from the other noise sources \citep{goncharov+2021b,Hazboun_2020,Lam_2016,Lam_2017,Lentati:2016ygu}. Intrinsic noise could come from jitter in the location of the emission mechanism, i.e. the beam could be non-stationary on the pulsar \citep{Lam_2016} or it could from various changes in the interior of the neutron star, effectively changing the moment of inertia in a stochastic manner \citep{melatos+2014,Cordes_2010}. Extrinsic noise could come from the intervening medium, i.e. the interstellar medium (ISM), from measurement noise, or from gravitational waves. In the case of X-ray observations, the ISM noise is essentially non-existent, since its effects decrease significantly with increases in the observation frequency \citep{Stinebring_2013}.  The Neutron Star Interior Composition Explorer (\nicer) gives us a unique chance to perform high-precision, long-term timing of pulsars in the X-ray band, free from ISM noise.  The detection of red noise that we document in this paper is significant because it is a detection of red noise where we {\em know} that none of the red noise is due to the ISM.  

In this paper we have applied existing techniques to X-ray data that traditionally have been used in radio data \citep[see][and references therein]{aab+20,Arzoumanian_2018,Hazboun_2020}. We use existing software including \enterprise (Enhanced Numerical Toolbox Enabling a Robust PulsaR Inference SuitE) \citep{ENTERPRISE}, and a Python timing package, \pint \citep[\pint Is Not TEMPO3,][]{pint}, to search for red noise in X-ray times-of-arrival (TOAs).

In \S\ref{sec:data}, we discuss how we process \nicer data, briefly reviewing the various filters we use to select events more likely to come from the pulsar. In \S\ref{sec:methods}, we introduce the various types of noise to illustrate why X-ray data is important, how they are accounted for or measured in our model, and how we construct our model.  In \S\ref{sec:searchfornoise}, we describe the Bayesian analysis of our datasets leading to our measurement of red noise using X-ray timing. In \S\ref{sec:future} we also discuss the analysis of an extension of the \nicer mission, as well as the prospects of future high sensitivity X-ray missions, using our newly developed data simulation tools. In \S\ref{sec:discussion} we discuss our results and summarize our investigation.

\section{Observations and Data Processing}\label{sec:data}

The primary instrument on \nicer is the X-ray Timing Instrument (XTI) with 52 co-aligned X-ray concentrators, each paired with an active detector that records event time and pulse height information for each detected photon \citep{Gendreau_2017}. 

In addition to the photons of interest, the detectors on NICER are sensitive to cosmic rays, trapped particles, optical light, and other radiation that can produce photon-like events.
We filter out spurious events and periods of high background using a variety of criteria, as explained in \citet{Deneva_2019}.

The measured event times have a precision of about 40 ns and are referenced to UTC with an accuracy better than 100 ns, after the fine timing bias calibrations are applied in the standard Level 1 processing. 

In our processing we construct one pulse time-of-arrival (TOA) per NICER 
\texttt{ObsID}, which means one TOA per UTC day, typically with hundreds to thousands of seconds of exposure. Our data for \psrj include $337$ TOAs taken between 2017 June 25 and 2020 November 12 with a mean TOA error of 11.4 $\mu$s and a mean observing cadence of 8.2 observations per month. 
\begin{figure}[h]
    \centering
        \includegraphics[width=0.95\textwidth]{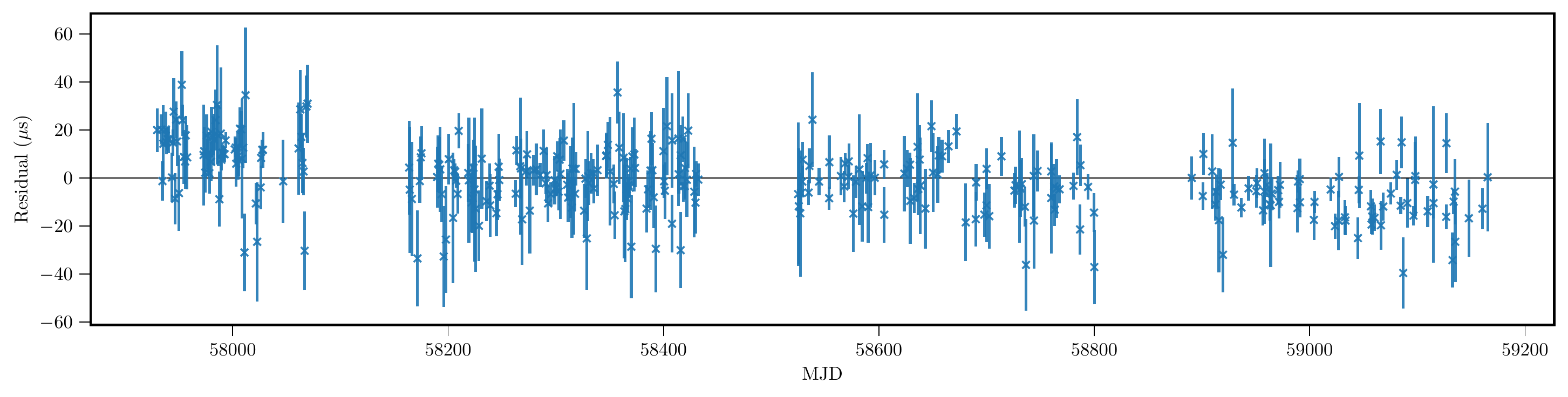}
        \caption{\nicer TOA residuals for \psrj using the \ftwo fit discussed in \S\ref{sec:searchfornoise}. \label{fig:j1824_resids}}
\end{figure}
Our data for \psrb include $466$ TOAs taken between 2017 June 28 and 2020 November 18 with a mean TOA error of 5.1 $\mu$s  and a mean observing cadence of 11.3 observations per month. 
\begin{figure}[h]
    \centering
        \includegraphics[width=0.95\textwidth]{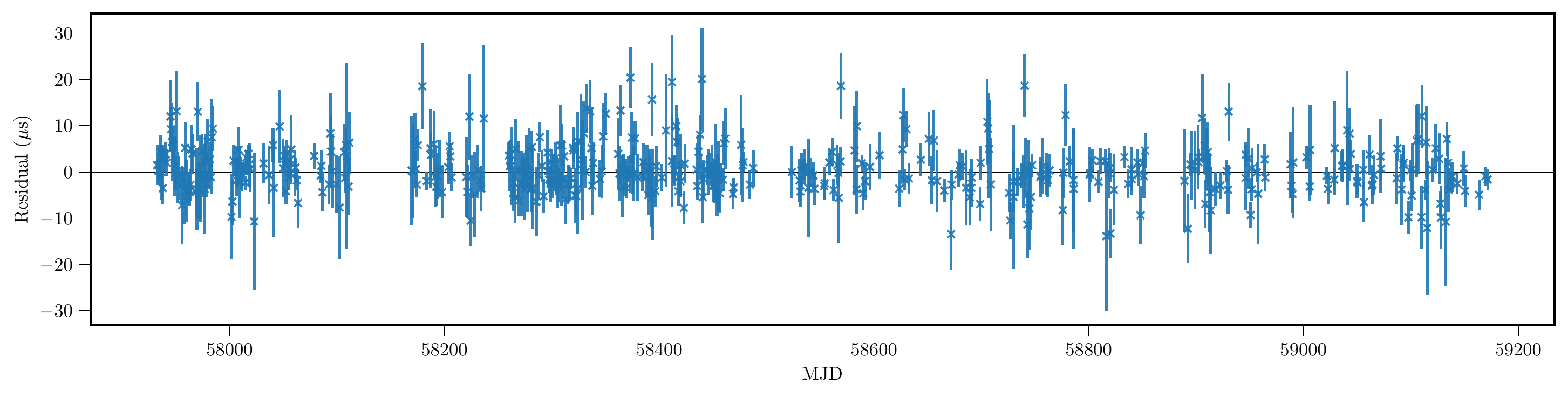}
       \caption{\nicer TOA residuals for \psrb using the \ftwo fit discussed in \S\ref{sec:searchfornoise}. \label{fig:b1937_resids}}
\end{figure}
Figures~\ref{fig:j1824_resids} and \ref{fig:b1937_resids} show the residuals (TOA $-$ timing model) and errors for \psrj and \psrb, respectively. In observations of \psrj we filtered out photons outside of the 1--5.5 keV range, while in  observations of \psrb we filtered out photons outside of the 1.15--5.55 keV range. TOAs for each of these pulsars are calculated using maximum likelihood fits to an analytic pulse template.
More complete details of the NICER timing accuracy and TOA generation procedure can be found in Section 2 of \citet{Deneva_2019}.

\section{Methods}
\label{sec:methods}

Understanding the different noise sources is crucial to our understanding of pulsars themselves and also the key to our ability to use them to detect gravitational waves \citep{Arzoumanian_2018}. 

We divide noise in pulsar timing residuals into two categories, white and red, which is a reference to the spectrum of the fluctuations.  The spectral density of white noise is independent of frequency, while red noise has larger amplitude at lower frequency.  To the eye, red noise in a time series appears as if the data wanders over long timescales.

We adopt our model for this work from PTAs, which use the data from many pulsars to observe the stochastic background of gravitational waves from supermassive binary black holes \citep{Lentati:2016ygu,IPTADR2}. This signal manifests in the timing data as red noise that is additionally spatially correlated, i.e., dependent upon the observation angle between pulsar line-of-sights from Earth \citep{HD}. The standard likelihood for gravitational-wave analysis with PTAs is well documented in the literature \citep{Lentati_2013,van_haasteren_2013,Demorest_2013,Lentati_2014,van_haasteren_2015,Arzoumanian_2016}. Here we forego the spatially correlated part of the signal model and search only for generic red noise within a single pulsar dataset. The covariance matrix is built from the various white noise components discussed in \S\ref{subsec:wn} below and the red noise is modeled using a Gaussian process with a Fourier basis, separately parametrized for each pulsar, as discussed in \S\ref{subsec:rn}. The model includes a linearized timing model where the timing model parameters are marginalized over during the analysis.

\subsection{White Noise}\label{subsec:wn}
Additional noise parameters are often used to augment pulsar timing data errors. Here we discuss EFAC and EQUAD\footnote{ECORR refers to noise which is correlated between TOAs within a particular observing epoch, across radio frequencies, but not correlated between epochs \citep{abb+14,Lam_2016}. Because ECORR is correlated across frequencies, and necessitates multiple TOAs from a single observing epoch, it is not required in our analysis.}, which were used in our analysis \citep[see][and references therein for further details.]{Lam_2016}
EQUAD, $Q$ is an error added in quadrature to TOA errors that models short timescale errors from diffractive scintillation, similar propagation effects, and pulse jitter. EFAC augments the TOA errors as a multiplicative factor, $F$, and models underestimates in TOA errors from low S/N ratio TOAs as well as allowing timing model fits to achieve reduced $\chi^2$-fits of one. 
Therefore the most generic model for the white noise in \nicer data is 
\begin{equation}
\sigma^2_{\rm Total}=F^2\sigma_{S/N}^2+Q^2 ,
\end{equation}
where $\sigma_{\rm Total}$ is the total white noise value used to construct the covariance matrix. We will see in \S\ref{subsec:wn_model} that the additional parameters $Q$ and $F$ were in fact not favored in a Bayesian model selection with \nicer data for \psrj.

\subsection{Red Noise}\label{subsec:rn}

Red noise, can be categorized into chromatic and achromatic processes, where {\it chromatic} here refers to a dependence on the frequency of the pulsed light from the neutron star.

In radio observations of pulsars the largest chromatic red noise comes from the bulk movement of the ISM across the line of sight to a pulsar which causes changes in the integrated electron column density, known as the dispersion measure (DM), and hence the dispersion of the pulses \citep{Shannon_2017,2018ApJS..235...37A}. DM effects scale as $1/\nu^2$ where $\nu$ is the electromagnetic frequency. Higher order effects (e.g. $1/\nu^4$) can also stem from scattering of the radio pulses through the ISM and secondary frequency dependent dispersion \citep{Hemberger_2008}. 

Achromatic red noise can be the result of many different processes. For example, magnetospheric state switching can affect the neutron star rotation itself \citep{Lyne_2010}.  Some MSPs can have asteroid belts which can cause perturbations in their orbits which appear as achromatic red noise in the residuals \citep{Shannon_Asteroid}.

Understanding the nature of red noise is crucial to building accurate pulsar timing models \citep{ColesandHobbs11,Stinebring_2013}.  Furthermore, gravitational wave detection using pulsars, i.e. pulling a very weak signal out of noisy data, requires an accurate description of the noise \citep{Lam_2016,Lam_2017,aab+20}. 

Various scaling laws, using optimal statistics to build signal-to-noise ratios, $\rho$, have been developed extensively in the literature for understanding when/if a pulsar data set is sufficient for the detection of red noise. In \S\ref{sec:future} we will use a single pulsar version of the $\rho$ expression in \citet{ccs+15},
\begin{equation}\label{eq:os_snr}
\rho = \left( 2T
\int^{f_H}_{f_L}  df
\frac{b^2f^{-2\gamma}}{\left(bf^{-\gamma}+2\sigma^2 \Delta t\right)^{2}}\right)^{1/2},
\end{equation}
for detecting power law red noise. Here $\gamma$ is the power law spectral index, $\Delta t$ is the time between observations (1/cadence), the length of the dataset is given by $T$, the white noise is given by $\sigma$, and where b is defined using both the amplitude and spectral index of the red noise as the following
\begin{equation}
b\equiv \frac{A_{\rm RN}^2}{12\pi^2}\left(\frac{1}{f_{\rm yr}}\right)^{3-\gamma}\,, 
\end{equation}
where $f_{\rm yr}$ is the reference frequency, here chosen to be 1/year.

X-ray observations can play a key role in the understanding of these types of noise for the following reason: there is no chromatic noise in X-ray observations. Radio waves are coherently scattered in the ISM with a strength depending on the inverse square of the frequency, strongly affecting the observed light through dispersive delays, scintillation, multipath propagation, etc.  X-rays are (photoelectrically) absorbed or incoherently (Compton) scattered, resulting only in simple, static attenuation of the light \citep{longair2011}. Therefore observing in X-ray data eliminates a significant category of noise. Further, in many pulsars the chromatic red noise measured is orders of magnitude larger than the various achromatic red noise types \citep{IPTADR2}. Finally mismodeled chromatic red noise can appear as achromatic noise \citep{Shannon_2016} so eliminating chromatic noise from the data set will allow for a proper characterization of achromatic noise. 

Since the achromatic red noise searched for in pulsar data is a stochastic signal, Gaussian-process regression \citep{rw06} is often used as the main tool for modeling. Here a normal kernel process was used in the Fourier domain with a power law prior  \begin{equation}
    P_{\rm RN}=\frac{A_{\rm RN}^2}{12\pi^2}\left(\frac{f}{f_{\rm yr}}\right)^{-\gamma} {\rm yr}^3 \label{eqn:achrom_rn}
\end{equation} for the power spectral density (PSD), a technique used often by PTAs \citep{van_haasteren_2013, Lentati:2016ygu}. There are $n$ frequencies considered ranging from $(1/T,n/T)$, where $T$ is the time span of the dataset. In longer datasets $n$ is usually $30$, but with these much shorter datasets we used $n=10$, since the red noise will only manifest in the lowest few frequencies. The PSD is given in units of TOA residual power, ${\rm yr^3}$. The amplitude, $A_{\rm RN}$ is unitless and referenced to a frequency of 1/year. The prior used in our search is log-uniform$(10^{-20},10^{-11})$. The spectral index, $\gamma$, prior was uniform$(0,7)$, where $\gamma=0$ is equivalent to white noise. Both are standard for pulsar noise searches \citep{aab+20,12yr_wideband}. 

In addition, we undertook a search for excess noise power more generically using a free spectral model. Such a search is unrestricted by any PSD model, and creates a posterior for the amplitudes for all frequencies searched, \citep{2018ApJS..235...37A}. The priors for these parameters were log-uniform in the range $(10^{-10},10^{-4} s)$. 

\subsection{Software}\label{sec:software}
The models were built using the \enterprise Python package developed for full pulsar timing array analyses \citep{ENTERPRISE}.  Two different techniques were used for sampling the likelihood. We used a standard Markov Chain Monte Carlo (MCMC) sampler, \ptmcmc, for doing parameter estimation and model selection, while a nested sampler, \texttt{Dynesty}, was used in order to calculate the evidence for the red noise likelihoods \citep{PTMCMC, Dynesty}. This evidence calculation allowed us to calculate the Bayes factor with respect to the base noise model, in this case a model that only includes the TOA errors. The purely Python based timing package, \pint was used to fit pulsar timing models to our TOAs, as well as for adding noise to our data simulations \citep{pint}. 

The noise was simulated using a set of routines built explicitly for adding noise into these X-ray datasets with \pint and is available on GitHub\footnote{\url{https://github.com/Hazboun6/pta_sim/blob/master/pta_sim/pint_sim.py}}. The routines are based on a set of publicly available routines in a Python wrapper package for the pulsar timing software \tempotwo \citep{hem06}, called \libstempo \citep{vallisneri:2015}. TOAs were zeroed out to match the deterministic timing model before adding noise. EFAC was added by inflating TOA errors before adding Gaussian-distributed noise, while EQUAD was added as a separate set of Gaussian-distributed perturbations. Red noise was added by pulling Fourier coefficients across thirty frequencies which were Gaussian-distributed around the chosen power law values.

Using this software we injected various known values of power law red noise and checked that our recovery with \enterprise was within expected ranges, given the sensitivity of the dataset. 

\section{Red Noise Search}
\label{sec:searchfornoise}

We undertook a series of analyses on the \nicer X-ray TOAs from \psrb and \psrj using \enterprise. Prior to these analyses the X-ray TOAs from each of these pulsars were fit in \pint using an ephemeris initially based on radio TOAs, but refit with \textit{only \nicer data}. This is an important milestone as both the \psrb and \psrj \nicer datasets are now mature enough to be independently fit. Following the analysis in \citet{Deneva_2019}, the TOAs were fit with the second derivative of the spin frequency, \ftwo, which effectively subtracts a cubic order polynomial fit from the data. A number of different physical processes can result in a significant detection of this parameter, including acceleration of the pulsar system, braking of the pulsar's spin from electromagnetic interaction with its surroundings, or the presence of a long period binary companion \citep[][and references therein]{Liu18}. 

\begin{figure}
    \centering
        \includegraphics[width=0.45\textwidth]{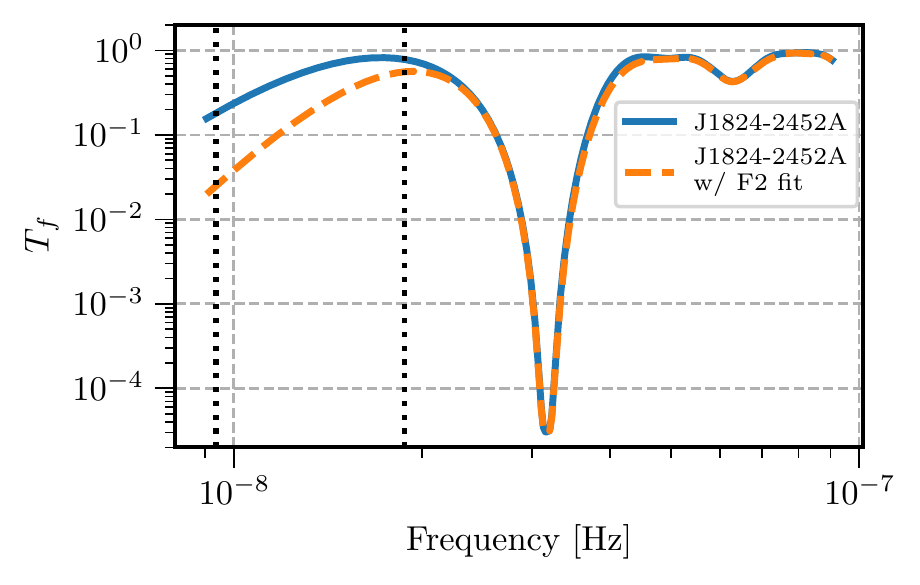}
        \caption{Transmission function for \psrj calculated for \nicer data using \hasasia \citep{hrs2019a} with and without a fit for \ftwo. $T_f$ shows the proportion of power transmitted through the timing model fit as a function of frequency. The two vertical dotted lines show $1/T$ and $2/T$, where $T$ is the time span of the dataset. The \ftwo fit removes a substantial amount of additional power at a frequency of the inverse time span of the data.\label{fig:F2fit}}
\end{figure}

A fit for \ftwo removes power from the data at frequencies proportional to the inverse time span of the data, see Figure~\ref{fig:F2fit}, and hence can absorb significant power from any red noise process \citep{Blandford1984, hrs2019b}. Just as in \citet{Deneva_2019} \ftwo was removed from the pulsar ephemeris model in the search for the red noise. This would leave power in the residuals that was initially removed by the \ftwo fit. Effectively this means that \ftwo was removed from the timing model design matrix. This fitting and then removing process ensures that any parameters covariant to \ftwo (such as $\dot{f}$, the quadratic spin down term) are within the linear regime for the red noise analysis, since a linearized timing model is used to marginalize over the timing model in our search. The parameters from these fits are presented in Table~\ref{table:wn_odds}. As a check the searches were also done with pulsar models that did not have a fit for \ftwo. These led to equivalent results, showing that the linear timing model marginalization was effective at bridging the gap for the purposes of these red noise analyses. Runs were also done where \ftwo was included in the design matrix. In the case of \psrj these showed much less support for red noise as expected.

Equation 2 in \citet{Johnston} allows for us to calculate our expected braking index given our fit for $f$, $\dot{f}$, and \ftwo for \psrj. This calculation yields a breaking index of around $2400$ which is not physically likely indicating that our observed \ftwo is not solely caused by the braking index.

\begin{table}
\begin{center}
\caption{Timing model for \psrj and \psrb.}
\begin{tabular}{l c c}
\hline
\hline
Parameter\tablenotemark{a} & \psrj & \psrb\\
\hline
Solar System Ephemeris\dotfill & DE438 & DE438\\
TT Realization\dotfill & TT(BIPM2019) & TT(BIPM2019)\\
Barycentric Time Scale\dotfill & TDB & TDB\\
Start (MJD)\dotfill & 57929.8 & 57932.5\\
Finish (MJD)\dotfill & 59165.3 & 59171.5\\
DILATEFREQ\dotfill & N & N\\
Number of TOAs\dotfill & 337 & 466\\
\textbf{Right ascension}\dotfill & \textbf{18:24:32.0077(2)} & \textbf{19:39:38.56133(6)}\\
\dotfill\textbf{(J2000) (hh:mm:ss.s)}&&\\
\textbf{Declination}\dotfill & {$\mathbf{-24:52:10.99(6)}$} & \textbf{21:34:59.126(2)}\\
\dotfill\textbf{(J2000) (dd:mm:ss)}&&\\
\textbf{Proper motion in}\dotfill & {$\mathbf{-0.7(6)}$} & {$\mathbf{-0.10(11)}$} \\
\dotfill\textbf{right ascension (mas/yr)}&&\\
\textbf{Proper motion in}\dotfill & \textbf{20(13)} & {$\mathbf{-0.51(17)}$}\\
\dotfill\textbf{declination (mas/yr)}&&\\
\textbf{Annual parallax (mas)}\dotfill & {$\mathbf{-0.6(14)}$} & \textbf{1.9(9)}\\
Epoch of position (MJD)\dotfill & 56999.9998000000000000 & 55321.0000000000000000\\
\textbf{F0 (s$^{-1}$)}\dotfill & \textbf{27.405534870665(6)} & \textbf{641.928221244395(4)}\\
\textbf{F1 (s$^{-2}$)}\dotfill & {$\mathbf{-1.735216(3) \times 10^{-13}}$} & {$\mathbf{-4.33071(2) \times 10^{-14}}$}\\
Epoch (MJD)\dotfill & 58547.5718813381904546 & 58552.0580235874009410\\
TZRMJD\dotfill & 56974.6086015081282523 & 55800.9121432622426041\\
TZRSITE\dotfill & ncyobs & gbt\\
TZRFRQ\dotfill & 1302.344971 & 812.187012\\
\hline
\end{tabular}
\end{center}
\tablenotetext{a}{Parameters in bold are those that were fit.}
\label{table:pars}
\end{table}

\subsection{White Noise Model Selection}\label{subsec:wn_model}

A hypermodel framework, \citep{hhh+2016}, was used to investigate whether any additional white noise parameters, as discussed in \ref{subsec:wn}, are needed in order to inflate the TOA errors of \nicer data. The hypermodel framework allows on-the-fly Bayesian model selection by using a hyper-likelihood built of the various models under investigation. These are combined, along with a parameter which chooses the specific model for which to evaluate the likelihood. This type of analysis was carried out recently in \citet{goncharov+2021a} for in-depth noise model selection on pulsars. An exhaustive analysis of models was done which included all combinations of EQUAD, EFAC and power law-modeled red noise for both pulsar datasets. The results for both pulsars are reported in Table~\ref{table:wn_odds}. The table gives the proportion of samples in a given model, normalized to the number of samples spent in the most favored model. The odds ratios can be read off by comparing two numbers in the same column, e.g., the odds ratio of Model~C to Model~D in \psrj is $1\!:\!0.77$. Entries of zero mean that a model was never visited. In the case of \psrj the preferred model was Model~C, {\it without any} additional white noise parameters, but {\it with} a power law red noise model. For \psrb the most preferred model was Model~B that included {\it only} EQUAD and {\it not} EFAC or red noise. In both pulsars models containing EFAC were highly disfavored \textemdash\ evidence that this parameter is not necessary for these X-ray datasets. Models containing EQUAD were only moderately favored or disfavored compared to models with red noise only, or red noise plus EQUAD. As the red noise becomes more distinguishable in these datasets the preference for EQUAD might become more clear.

\begin{table}
\centering
\begin{tabular}{c|c|c|c|c|c}
\multicolumn{4}{c}{} &\multicolumn{2}{c}{Sample Fraction} \\
Model & RN & EFAC & EQUAD & J1824$-$2452A & B1937+21   \\
\hline
A&&$\checkmark$&&0&0.01\\
B&&&\checkmark&$9\times10^{-5}$&\textbf{1}\\
C&\checkmark&&&\textbf{1}&0.54\\
D&\checkmark&&\checkmark&0.77&0.73\\
E&&\checkmark&\checkmark&0&0.02\\
F&\checkmark&\checkmark&&0.06&0.01\\
G&\checkmark&\checkmark&\checkmark&0.05&0.02\\
\end{tabular}

\caption{Sample fractions for the noise model selection in \psrj and \psrb. A check mark designates that a given element was included in the model. The fractions are normalized by the most favored model, bolded for the two pulsars. The odds ratio can be read off by comparing two number from the same column, i.e. the odds ratio of Model~B to Model~C for \psrb is $1\!:\!0.54$.}
\label{table:wn_odds}
\end{table}

\subsection{Red Noise Analysis and Evidence Calculation}

\begin{figure}
    \centering
        \includegraphics[width=0.45\textwidth]{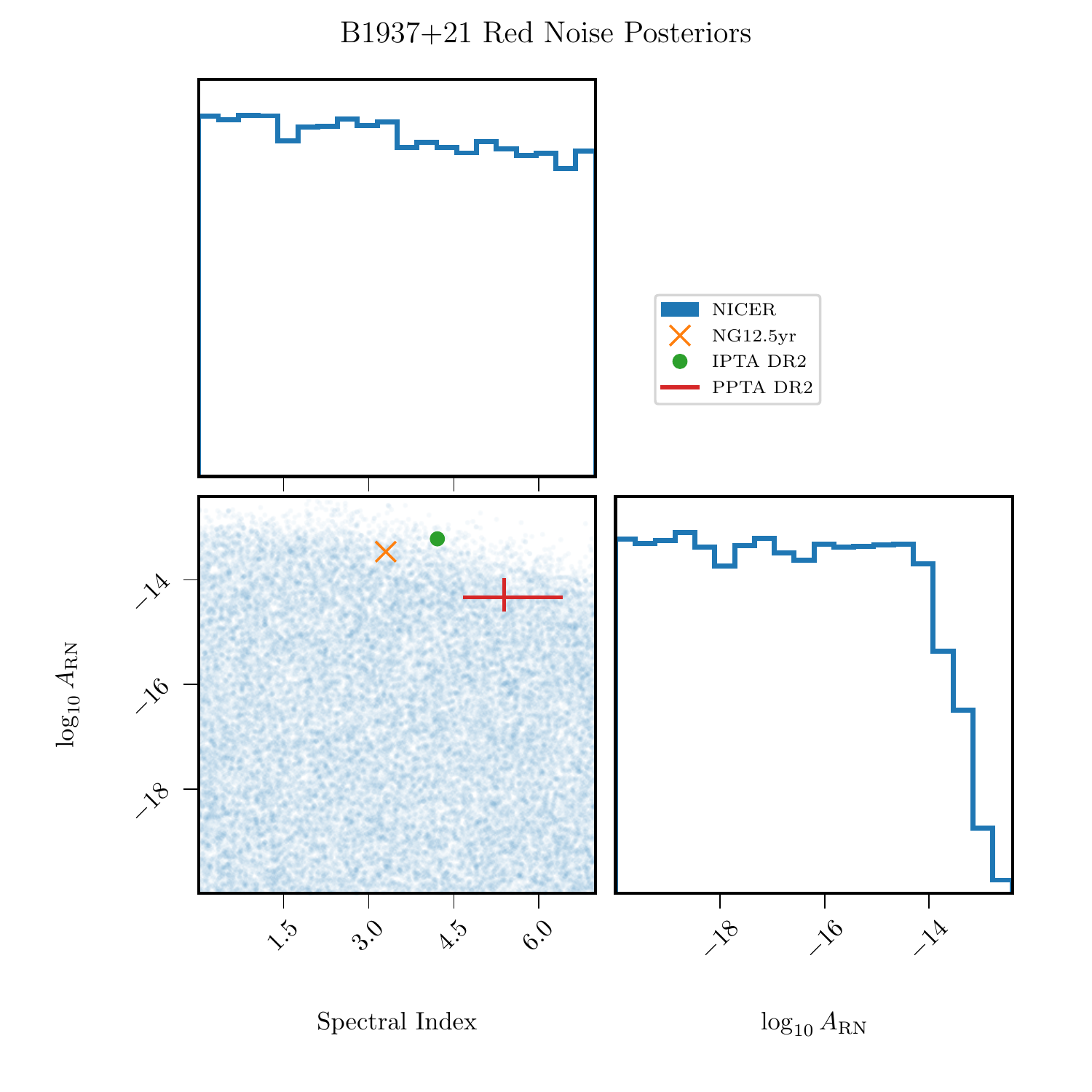}
        \caption{Bayesian 2-dimensional posterior for power law red noise model in the \psrb dataset. The X-ray data is currently fairly uninformative, however the upper extent of the posteriors is nearing the measured values in \citep{aab+20,goncharov+2021a,IPTADR2}. \label{fig:real_b1937_plaw}}
\end{figure}

\begin{figure}
    \centering
        \includegraphics[width=0.45\textwidth]{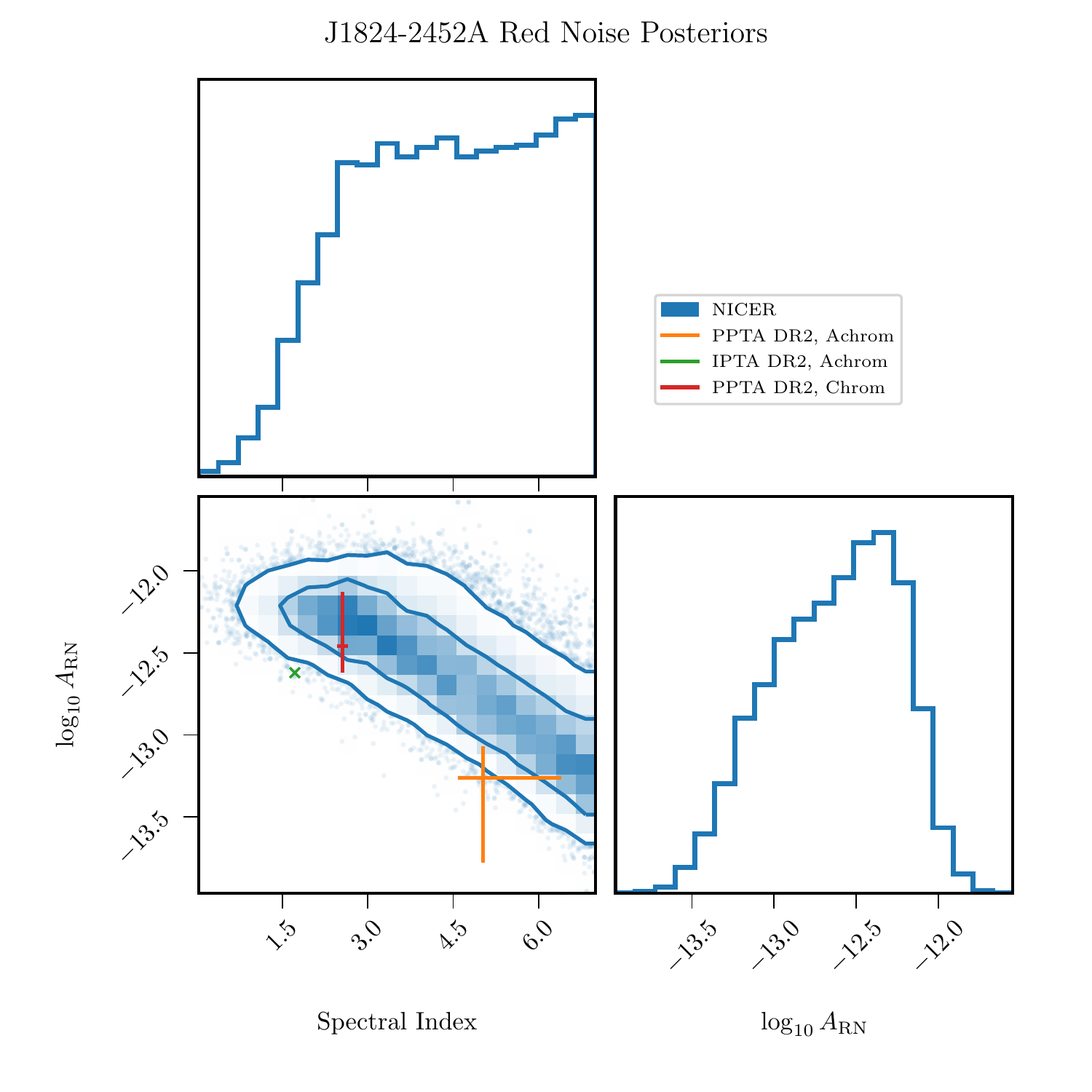}
        \caption{Bayesian 2-dimensional posterior for power law red noise model in the \psrj dataset. While the amplitude posterior from the X-ray data shows a significant detection, the spectral index is fairly unconstrained. Values from other PTA datasets \citep{aab+20,goncharov+2021a,IPTADR2} are also shown, along with errors when available. Note that the spectral index/ amplitudes for the PPTA chromatic and achromatic process are both in fair agreement with the \nicer posterior. } 
    \label{fig:real_j1824_plaw}
\end{figure}

We proceeded to run the analyses in \psrb and \psrj with the most preferred model for each pulsar found in \S\ref{subsec:wn_model}.  In addition to the power law model for the PSD of the red noise used above, we also used a free spectral model where the amplitude of the PSD at each of ten frequencies are free parameters and are not restricted to a power law model.  The parameters for the power law search are shown in Figures~\ref{fig:real_b1937_plaw} and \ref{fig:real_j1824_plaw} and are compared to red noise values from longer time span radio timing data. In the case of \psrb the posteriors are not very informative, except to set an upper limit on the noise in the X-ray data. Note that the radio data parameters are close to the sensitivity threshold of the \nicer dataset. The power law red noise posterior for \psrj shows a significant detection of red noise that is in broad agreement with various radio data results. 

In order to quantify the significance of both of these models Bayesian evidence calculations were carried out using the nested sampling Python package \texttt{Dynesty}. This was done for models that included power law red noise models and were compared against the models without red noise. The most favored models from the model selection analysis, see Table~\ref{table:wn_odds}, were used as the base models. In the case of \psrj, since red noise was the only preferred additional noise element, the noise-only model included TOA errors plus a linear timing model perturbation. In the case of \psrb the base model had TOA errors, a linear timing model perturbation and EQUAD. The Bayes factors calculated are given in Table~\ref{table:bf}. 

\begin{table}
\centering
\begin{tabular}{c|c}
PSR 				&	$\ln (BF)$ \\
\hline
B1937+21			&	$-0.31 \pm 0.01$ \\
J1824$-$2452A	&	$9.634 \pm 0.016$ \\
\end{tabular}

\caption{Bayes factors for the presence of red noise in \nicer data. The data for \psrj shows a clear detection of achromatic red noise, while the data for \psrb shows no strong evidence for the presence or absence of red noise. }
\label{table:bf}
\end{table}
We find a Bayes factor very close to one for the red noise search in \psrb, therefore we cannot claim or refute the presence of red noise in the \nicer data. We find strong evidence for a power law red noise model in \psrj with a natural log Bayes factor of $9.634 \pm 0.016$.
\begin{figure}
    \centering
        \includegraphics[width=0.45\textwidth]{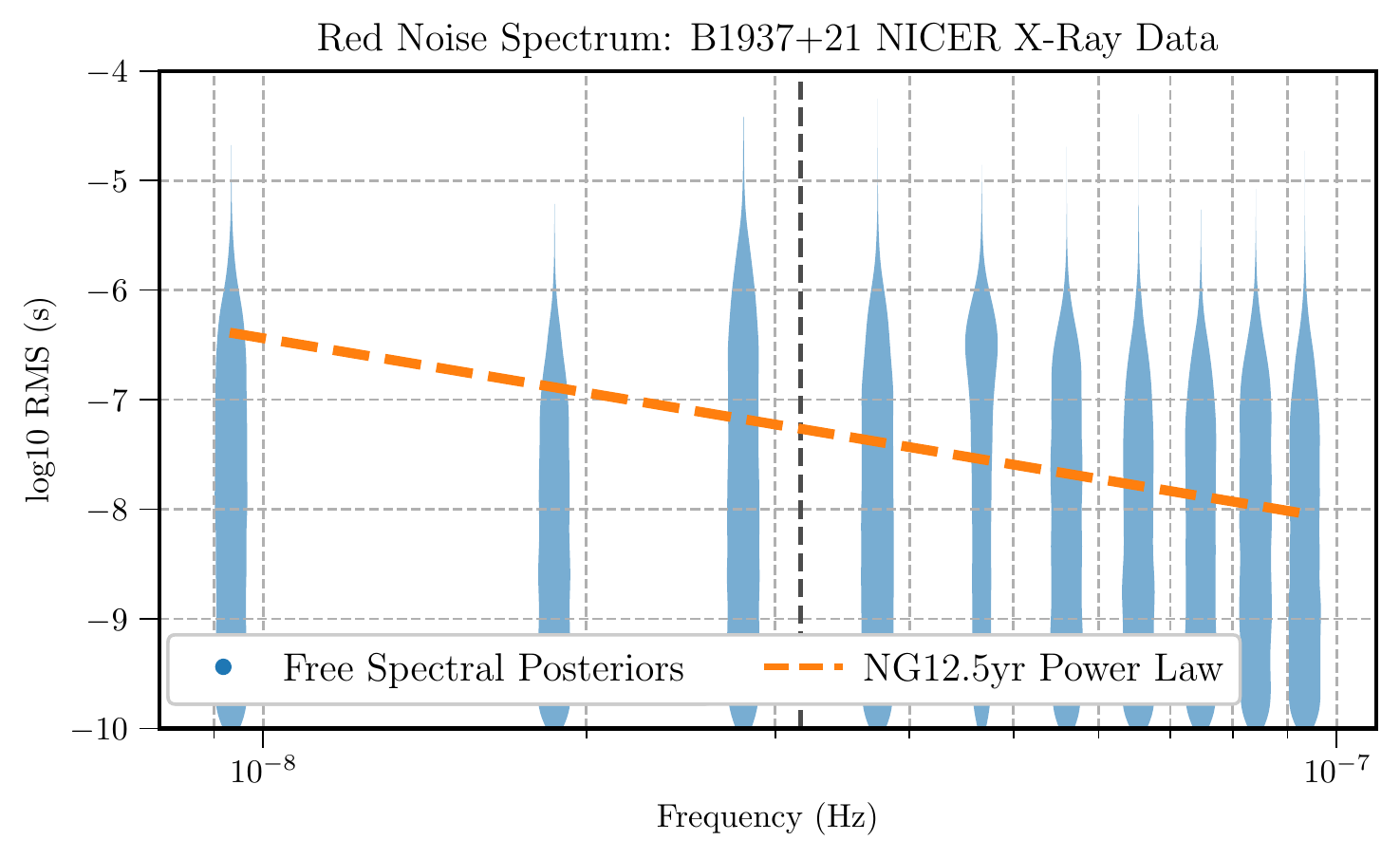}
        \caption{Free spectral analysis posteriors from \psrb \nicer data. Each of the violin plots shows the posterior probability that parameterizes the amplitude of noise in the respective frequency bin. The dashed orange line shows the best fit power law noise from \citet{aab+20}. \label{fig:real_b1937_fs}}
\end{figure} 

\begin{figure}
    \centering
        \includegraphics[width=0.45\textwidth]{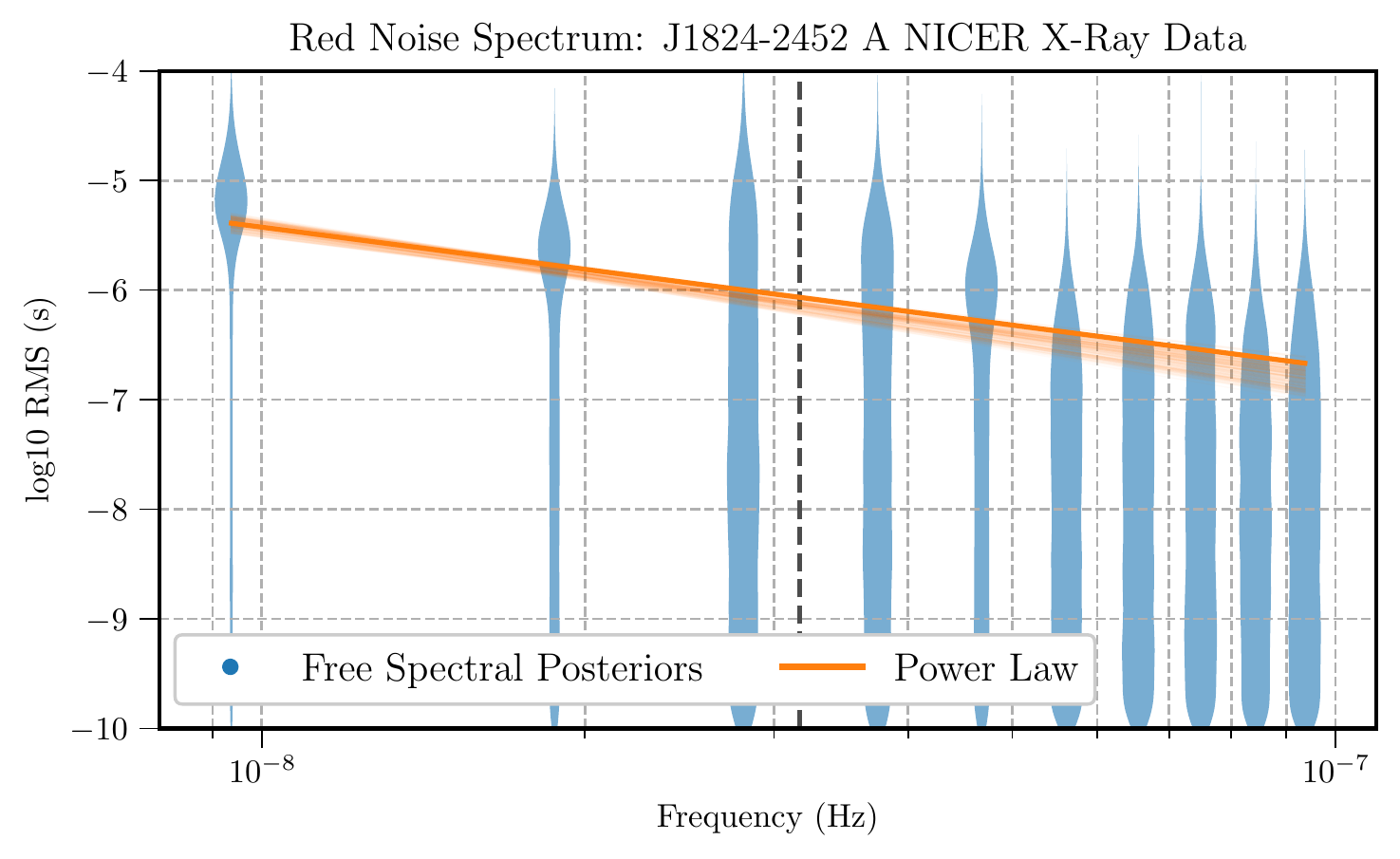}
        \caption{Free spectral analysis posteriors from \psrj \nicer data. Each of the violin plots shows the posterior probability that parameterizes the amplitude of noise in the respective frequency bin. The thin tails extending to the minimum ${\rm RMS}$ values in the two lowest frequency bins represent significant detections of power in those bins. The solid orange lines show the 2 dimensional maximum {\it a posteri} value (bold) and a number of other realizations of the power law from the X-ray data analysis.}
    \label{fig:real_j1824_fs}
\end{figure} 

\begin{figure}
    \centering
        \includegraphics[width=0.45\textwidth]{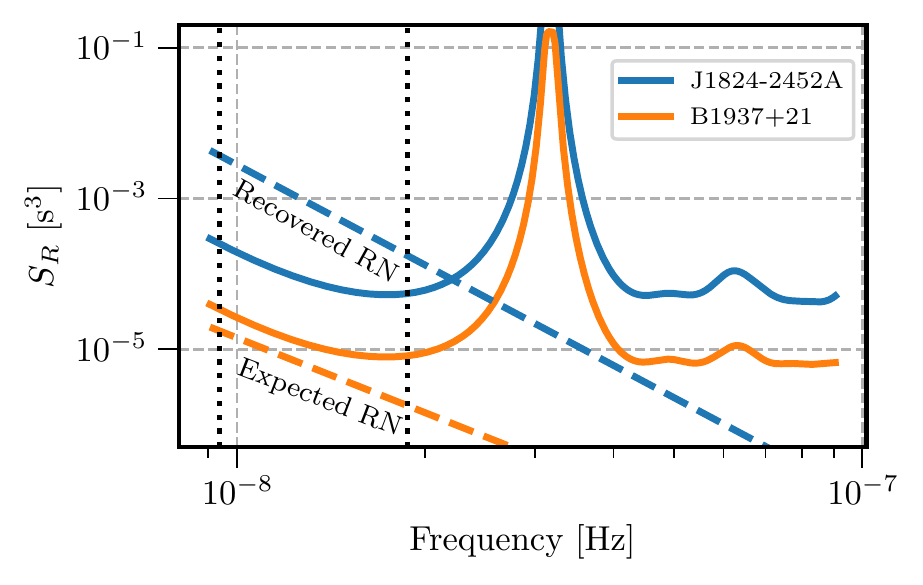}
        \caption{Comparison of pulsar sensitivity curves with power law red noise. The solid curves show the PSD of the residuals for the two \nicer datasets. The dashed lines show the best fit power law models for red noise from our analysis  for \psrj and from \citet{aab+20} for \psrb. The red noise power is larger than $S_R$ for the lowest two frequencies in \psrj, but does not rise above the curve even in the lowest frequency of \psrb.}
    \label{fig:senstitivity_comparison}
\end{figure}

The theoretical noise PSD, $S_R$, often referred to as the sensitivity curve or just the sensitivity, of individual pulsars is well understood in the context of pulsar timing model fits and various sources of white noise \citep{hrs2019b}. These datasets are strongly affected by the short time span, especially when looking for steep red noise. The fit for the spin down parameters pulls power out at the lowest frequencies important for the detection of red noise, while the astrometric fit leaves a broad peak at 1/year \citep{Blandford1984}. Figure~\ref{fig:senstitivity_comparison} shows the sensitivities of both of these pulsar datasets made using \texttt{hasasia} \citep{hrs2019a}, along with red noise PSDs drawn with the parameters retrieved  in this analysis (in the case of \psrj) or from longer radio datasets (in the case of \psrb.) It is obvious from the comparisons that we would expect to detect the much larger amplitude process in \psrj, but not detect the red noise in \psrb, since the radio timing parameters put the red noise below the sensitivity of this pulsar.

\section{Future Prospects: Simulated X-ray Data}\label{sec:future}

Future X-ray data sets may well be able to detect red noise in \psrb, and other pulsars, as we did in \psrj. We ran simulations to determine what specifications (cadence and TOA error) would allow a mission to make such a detection. These simulations allow us to assess when a future X-ray mission, like the proposed Spectroscopic Time-Resolving Observatory for Broadband Energy X-rays (STROBE-X), will detect red noise in \psrb using X-ray data alone.

\subsection{Simulations}\label{subsec:datasim}

To simulate residuals, we extended the \nicer 3.5 year dataset that was integrated by ObsID (data that was collected on the same UTC day) forward in time, and matched the current observing cadence as closely as possible using the following algorithm. First, we took the difference between the Modified Julian Dates (MJDs) of two adjacent TOAs chosen at random within the \nicer dataset. We added this difference to the MJD of the most recent TOA in the growing dataset to create the MJD for the next TOA. We repeated this process until the dataset created after the end of \nicer data reaches our desired length (e.g. 5 years). The result created was an extended dataset with similar observing cadence but non-identical spacing in time. We also scaled the observing cadence to create data with varying cadences. This allowed us to create datasets with different observing cadences from the \nicer data but with realistic variation. We simulated errors on these new points by taking the error of a randomly chosen point in the actual \nicer data and making that the error of a TOA in the simulated dataset. We also scaled these errors by a multiplicative factor determined by the ratio between the desired error and the mean error of the \nicer data in order to test data with smaller or larger error values. Once we have created the simulated TOAs, we use the \pint scripts mentioned in \S\ref{sec:software} to subtract the calculated residual from each TOA so that the simulated TOAs fit the timing model.

\subsection{Noise Models in \psrb}\label{subsec:wn_favor}

The most conservative model for detecting red noise in \psrb is one that has red noise, EQUAD, and EFAC all injected, as that adds the most noise into the data. Conversely, the most optimistic model is a model that only has red noise without EQUAD or EFAC. For the purposes of determining if red noise will be detected in \psrb at a given observing cadence, TOA error, and mission length, any other noise model would fall between these two models in terms of our ability to make a detection of red noise. Given the expensive nature of these simulations, these two cases were adopted to bracket our understanding of \nicer's ability to detect red noise in \psrb in the near future. 

\subsection{Detectability of Red Noise in Future X-ray Data}
\begin{figure}
    \centering
    \includegraphics[scale=.8]{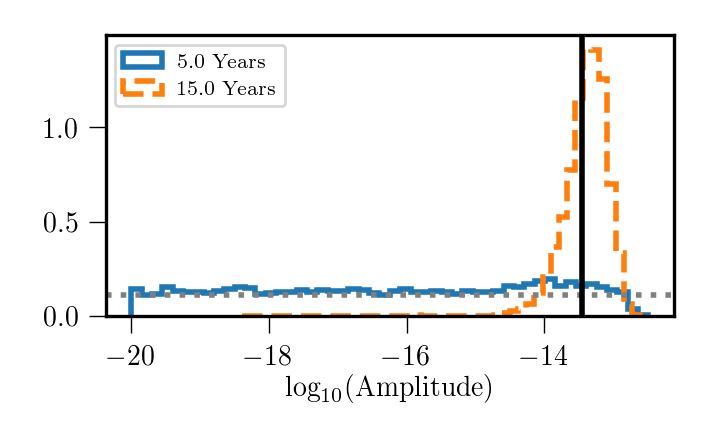}
    \caption{Posterior of recovered $\log_{10}A_{\rm RN}$ values for 5 years of simulated \nicer data on \psrb with a cadence of 10 observations per month and an error value of 5 microseconds (solid blue). This is close to the real \nicer data cadence and error value. Posterior of recovered amplitude values for 10 years of simulated \nicer data on \psrb (dashed orange). The dotted gray line across the plot represents the priors and the vertical black line represents the injected log of the red noise amplitude of $\log_{10}A_{\rm RN}=-13.46$. At 5 years there is no detection of red noise, though we can see an upper limit beginning to be set on the posterior. At 15 years there is a clear detection of the injected red noise.}
    \label{fig:HistogramAmpDetection_FakeData}
\end{figure}

We injected red noise into the simulated data sets at the level currently observed in the NANOGrav 12.5-year dataset ($\gamma=3.387$ and $\log_{10}A_{\rm RN}=-13.46$) \citep{aab+20}. We conducted the analysis in the same manner as for the existing \nicer data, described in \S\ref{sec:software}, this time searching for a signal that we had injected. Figure \ref{fig:HistogramAmpDetection_FakeData} shows the posterior probabilities for red noise amplitude recovered in our analysis of simulated \nicer data for two different mission lengths (5 years as the solid blue line, and 10 years as the dashed orange line). The dotted gray line across the plot shows the prior, while the black vertical line shows the injected value.

In Figure~\ref{fig:HistogramAmpDetection_FakeData} we see the difference in the character of the posteriors for an upper limit (solid) and a detection (dashed), and also the difference that a dataset three times as long makes in our ability to detect red noise.  The 15-year dataset has a narrow posterior, tightly localized around the injected red noise value of $\log_{10}A_{\rm RN}=-13.46$. Hence we were able to successfully recover the injected value. By contrast the 5-year posterior resembles the prior except in the high amplitude region that is ruled out by the data.  While we are able to set an upper limit, this particular analysis is not a detection.

\begin{figure}
    \centering
    \includegraphics[scale=.8]{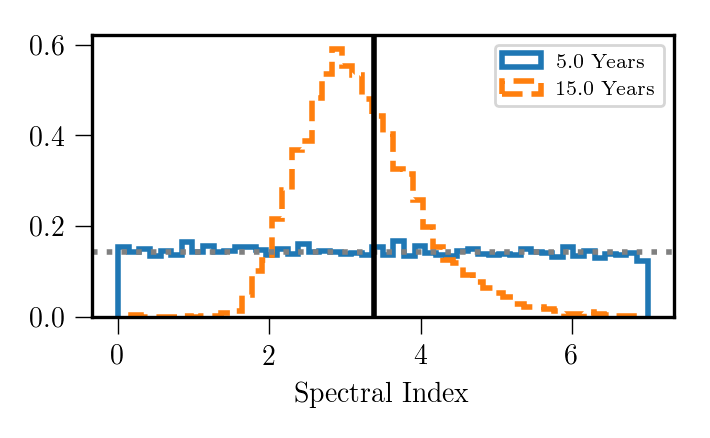}
    \caption{Posterior of recovered spectral indices for 5 years of simulated \nicer data on \psrb (solid blue) and for 15 years of simulated \nicer data on \psrb (dashed orange). The horizontal dotted gray line across the plot represents the priors and the vertical black line represents the injected spectral index of $\gamma=3.387$. At 15 years the posterior is localized, indicating a more significant detection, whereas at 5 years the posterior is similar to the priors.}
    \label{fig:HistogramSpectralIndDetection_FakeData}
\end{figure}

Figure \ref{fig:HistogramSpectralIndDetection_FakeData} shows the posterior probabilities for the power law spectral index recovered in our analysis of simulated \nicer data of two different lengths (5 years as the solid blue line, and 15 years as the dashed orange line). The dotted gray line across the plot shows the prior, while the black vertical line shows the injected value. 

Again, we see that the longer dataset in Figure \ref{fig:HistogramSpectralIndDetection_FakeData}, has a clear high probability density region in the posterior, with the most likely value settling in a range near $\gamma=3$. This is near the injected spectral index of $\gamma=3.387$.  By contrast in the analysis of the shorter dataset the posterior still closely resembles the priors.

\subsection{Detecting red noise in \psrb}

\begin{figure}
    \centering
    \includegraphics[scale=.8]{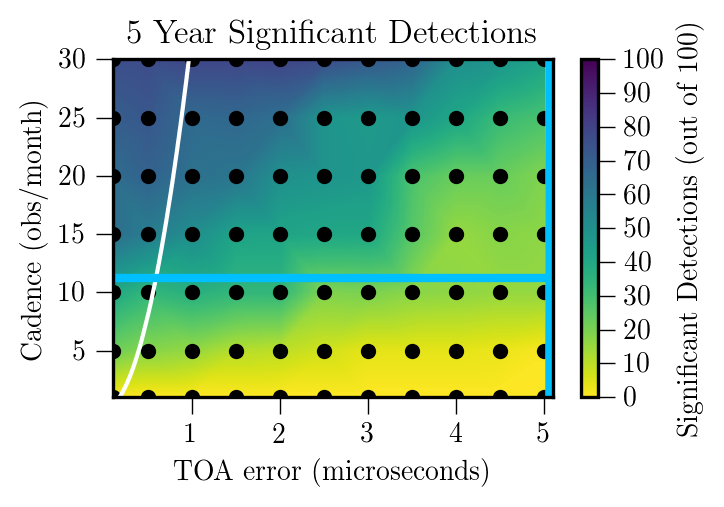}
    \includegraphics[scale=.8]{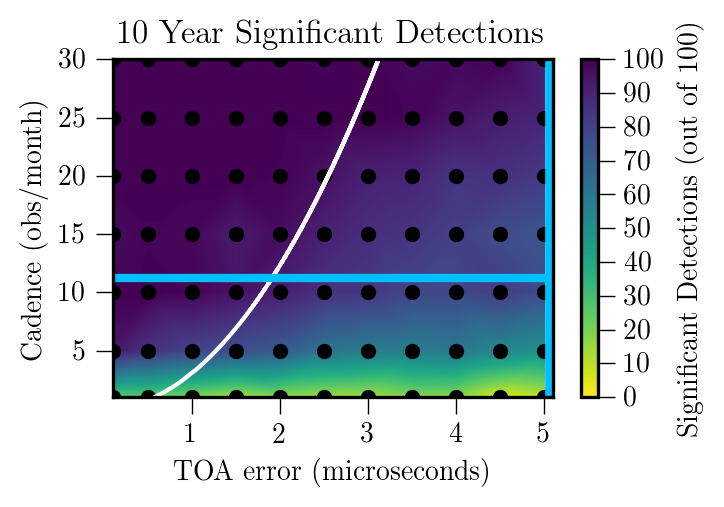}
    \caption{Number of significant red noise recoveries as a function of TOA error and cadence for a 5-year mission (left panel) and a 10-year mission (right panel) with injected values of $\log_{10}A_{\rm RN}=-13.46$ and $\gamma=3.387$. The noise model was used on simulated data constructed from \nicer data following the procedure in \S\ref{subsec:datasim}. Each point (shown in black) was constructed using at least 100 different iterations of simulated data and noise injected following the procedures in \S\ref{subsec:wn} and \S\ref{subsec:rn} using a model with red noise, EQUAD, and EFAC. The color of the graph, interpolated from nearby points, indicates how many of those simulations resulted in a Bayes factor larger than 100 and a recovered $\log_{10}A_{\rm RN}$ within 1 of the injected value. The light blue lines indicate the existing TOA error and cadence for the 3.5 year \nicer data. The white line indicates where there is a S/N of 3 according to equation \ref{eq:os_snr}. These figures show what the mission requirements for the next X-ray mission (such as STROBE-X) would need to be in order to detect red noise in \psrb (see text).}
    \label{fig:multicolor_extravaganza_WN}
\end{figure}

Figure~\ref{fig:multicolor_extravaganza_WN} reveals the strategies that a future 5-year X-ray mission (like STROBE-X) could use for detecting red noise in \psrb. In particular it shows the direct exchange that such a mission could take between TOA error and observing cadence and still detect red noise. For example, observations that yield 2-microsecond TOA error could be made 20 times per month and detect red noise.  Alternatively, 3-microsecond TOA error would require observations more frequently at 30 observations per month.

In order to achieve smaller uncertainty in TOAs than \nicer, STROBE-X would either need to have larger effective collecting area than \nicer, observe each pulsar for longer, or both.  For example, \nicer is achieving roughly 5 microsecond precision on \psrb using OBSID TOAs. In order to achieve 2.5 microsecond precision, STROBE-X would need to have 4 times the effective collecting area of \nicer, i.e. 7600 cm$^2$ instead of 1900 cm$^2$ or observe for 4 times as long (\nicer observes each TOA for hundreds or even thousands of seconds) \citep{Deneva_2019,mission_guide}.

For a longer 10-year X-ray mission, Figure \ref{fig:multicolor_extravaganza_WN} shows that red noise in \psrb will be widely detectable at a variety of observing cadences and TOA errors. Specifically, if \nicer continues to observe \psrb for a total of 10 years, it would be expected to detect red noise in the pulsar. This is shown in the region of the graph on Figure \ref{fig:multicolor_extravaganza_WN} closest to the intersection of the black lines. These represent current \nicer cadence and TOA error and the purple color is indicating a high likelihood of significant detection of red noise.

\begin{figure}
    \centering
    \includegraphics[scale=.8]{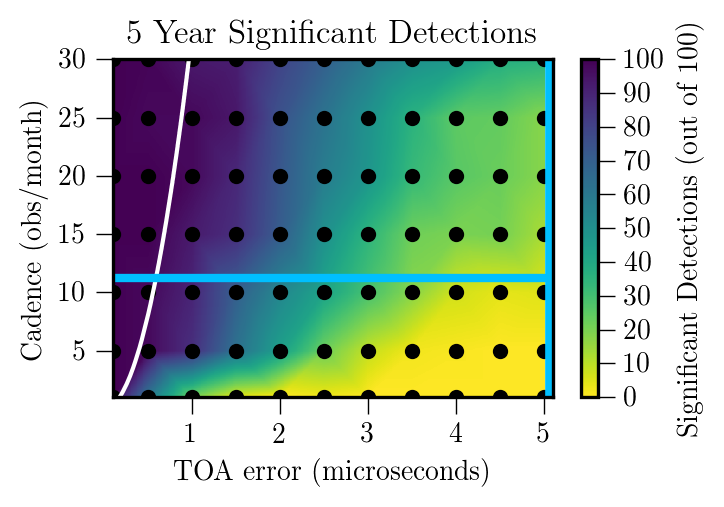}
    \caption{Number of significant red noise recoveries as a function of TOA error and cadence for a 5-year mission with injected values of $\log_{10}A_{\rm RN}=-13.46$ and $\gamma=3.387$. The noise fitting model was used on simulated data constructed from \nicer data following the procedure in \S\ref{subsec:datasim}. Each point (shown in black) was constructed using at least 100 different iterations of simulated data and noise injected following the procedure in \S\ref{subsec:rn} using a model with only red noise. The color of the graph, interpolated from nearby points, indicates how many of those simulations resulted in a Bayes factor larger than 100 and a recovered $\log_{10}A_{\rm RN}$ within 1 of the injected value. The light blue lines indicate the existing TOA error and cadence for the 3.5 year \nicer data. The white line indicates where there is a S/N of 3 according to equation \ref{eq:os_snr}.}
    \label{fig:multicolor_extravaganza_EFAC}
\end{figure}

While the simulated dataset used to create Figure \ref{fig:multicolor_extravaganza_WN} is injected with both EQUAD and EFAC according to \S\ref{subsec:wn}, the simulated dataset used to create Figure \ref{fig:multicolor_extravaganza_EFAC} only has white noise commensurate with the TOA errors and no EQUAD or EFAC. Accordingly, Figure \ref{fig:multicolor_extravaganza_WN} shows a more conservative analysis as it contains a larger amount of noise than \ref{fig:multicolor_extravaganza_EFAC}. Therefore, the likelihood of making a significant detection with the same cadence and TOA error is higher in Figure \ref{fig:multicolor_extravaganza_EFAC} than in \ref{fig:multicolor_extravaganza_WN}.

In order to compare our numerical simulations with the theoretical signal-to-noise ratio given in Equation~\ref{eq:os_snr} we need to relate $\rho$ to the Bayes factors we have been calculating. The Laplace approximation \citep{Romano,MacKay:2002},
$2\ln{}\mathcal{B} \approx \rho^2 + 2\ln\Big(\frac{\Delta{}V_1/V_1}{\Delta{}V_0/V_0}\Big)\,$, 
where $\Delta{}V_\mathcal{M}$ is the characteristic spread of the likelihood around the maximum, and $V_\mathcal{M}$ is the total parameter space volume, serves as a crude relation between the Bayes factor and the signal-to-noise ratio. The second term on the right-hand side is a negative term that encodes the Occam penalty for the use of too many parameters. A rough approximation can be made by assuming the second term is negligible. Our detection threshold of a Bayes factor of 100 then corresponds to $\rho\approx3$.

The white lines in Figures \ref{fig:multicolor_extravaganza_WN}, \ref{fig:multicolor_extravaganza_EFAC}, and \ref{fig:multicolor_extravaganza_GWB} show where $\rho=3$ for the injected red noise parameters. We would expect that along the white line there is a 50\% chance of detecting the injected red noise or GWB. Accordingly, the left panel of Figure \ref{fig:multicolor_extravaganza_WN} is more pessimistic than equation \ref{eq:os_snr} would indicate while the right panel of Figure \ref{fig:multicolor_extravaganza_WN} and Figures \ref{fig:multicolor_extravaganza_EFAC} are more optimistic.

\subsection{Detecting the Gravitational Wave Background}

\begin{figure}
    \centering
    \includegraphics[scale=.8]{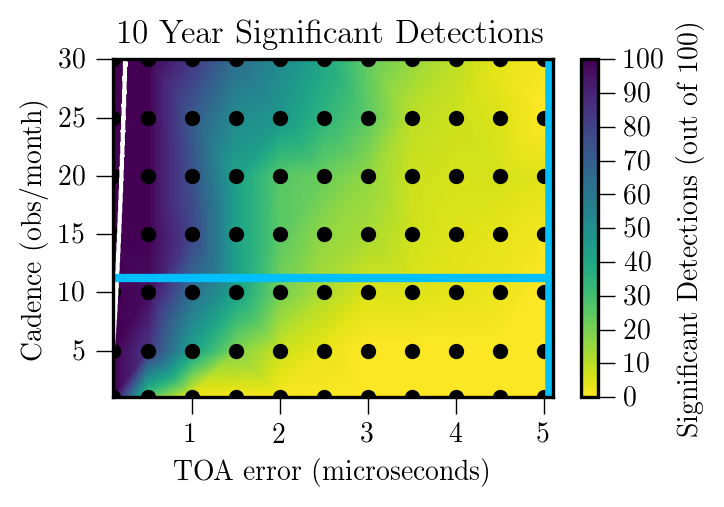}
    \caption{Number of times that the injected GWB was recovered as a function of TOA error and cadence for a 10-year mission with injected values of $\log_{10}A_{\rm GWB}=-14.699$ and $\gamma_{\rm GWB}=4.333$. The noise fitting model was used on simulated data constructed from \nicer data following the procedure in \S\ref{subsec:datasim}. Each point (shown in black) was constructed using at least 100 different iterations of simulated data and noise injected following the procedure \S\ref{subsec:rn} with a model that only includes red noise. The color of the graph, interpolated from nearby points, indicates how many of those simulations resulted in a Bayes factor larger than 100 and a recovered $\log_{10}A_{\rm RN}$ within $1$ of the injected value. The light blue lines indicate the existing TOA error and cadence for the 3.5 year \nicer data. The white line indicates where there is a S/N of $3$ according to equation \ref{eq:os_snr}. For this graph, the GWB was injected instead of the red noise and no additional white noise was injected. The red noise will overpower the GWB in \psrb in this time frame so this is not a realistic depiction of how long it will take to detect GWB in \psrb.}
    \label{fig:multicolor_extravaganza_GWB}
\end{figure}

We used our existing infrastructure to ask a similar question about the stochastic GWB rather than about a generic source of red noise.  Using the latest values from \citet{Arzoumanian_2018} we injected $\log_{10}A_{\rm GWB}=-14.699$ and $\gamma_{\rm GWB}=4.333$ into the simulated data for \psrb in place of red noise without injecting EQUAD or EFAC with a simulated 10-year mission length. The results are shown in Figure \ref{fig:multicolor_extravaganza_GWB}.  It is important to note that the GWB was injected in place of the red noise, not in addition to it, and that no additional white noise was injected. Because the red noise in \psrb is strong and our best model for \psrb shown in section \S\ref{subsec:wn_model} does include EQUAD, this is not a realistic representation of GWB detection in actual \psrb data. However, this analysis is useful in determining mission requirements for detecting GWB in a pulsar like \psrb that only has the GWB as noise in addition to the \nicer TOA errors. This would be the most optimistic scenario for the detection of the GWB in a pulsar using X-ray data. In such a pulsar, a future X-ray mission would need to have an observing cadence close to once a day with a TOA error under 2.5 microseconds, or a TOA error under 1.0 microseconds and a more infrequent observing cadence to detect the GWB with a 10-year mission. 

\section{Discussion and Summary}\label{sec:discussion}
Using NICER X-ray data alone, we detect red noise in \psrj with $\log_{10}A_{\rm RN}=-12.60^{+0.36}_{-0.46}$ and a spectral index of $\gamma=4.41^{+1.81}_{-1.83}$, in agreement with radio observations. X-ray observations are free from the influence of ISM that impacts radio data making this detection significant as it is a detection of red noise that is known not to be caused by the ISM. While the red noise in \psrb was below the threshold of a detection, our analysis of simulated future data will help design future X-ray missions and inform when such a detection will take place. The detection of achromatic red noise in \psrb is well known in the pulsar timing community \citep{ktr94, Lentati:2016ygu,aab+20,12yr_wideband}. As the \nicer dataset becomes more sensitive in the next few years it will be extremely interesting to see how much red noise is detected, and how much of the red noise seen in radio timing data is achromatic. These studies will be bolstered by the substantial amount of data from gamma ray timing of pulsars \citep{kerr+2015} by the Fermi Gamma-ray Telescope, and the ongoing effort to search for the GWB in Fermi data (Kerr, et al., Submitted). 

As has been the case in the pulsar timing community for the last decade, an analysis of noise models which yields the PSD of the noise, such as we have done, here replaces the $\sigma_z$ analysis outlined in \citet{Matsakis} and earlier attempts at using 2nd order structure functions \citep{cordes+1985} that were previously developed to quantify the stability of a pulsar using time domain methods. The clock community and the pulsar timing community are both interested in the stability of pulsars, and they each have different language to describe them, but we believe this analysis will appeal to both.  Both communities speak the language of PSD, and ultimately that is what both communities would like to describe their clocks, i.e. what is the power spectral density of the noise?

If one wishes to relate the PSD, e.g., our results, to $\sigma_z$, \citet{Matsakis} gives a relationship between the spectral index of a power law PSD in the residuals, $S_{\rm R}\propto f^{-\gamma}$, $0<\gamma<6$ and their statistic, $\sigma_z^2\propto\tau^{\gamma-3}$. Our results for \psrj of $\gamma\sim3$ would give a flat dependence on $\tau$, similar to the Allan deviation. Allan statistics have an exact correspondence to PSD power law models for spectral index, $0<\gamma<4$; $\gamma=3$ is formally a flicker-FM noise type. A low frequency cutoff filter can be employed, if necessary for the case $2<\gamma<6$ \citep{makdissi:2009}. Future analysis will implement Allan statistical treatments so that standard clock characterizations of \nicer data can be readily used by clock analysts \citep{howe:2006,howe+2021}

Future projects will improve on the \nicer dataset both in sensitivity and quantity of data.  We have used our analysis of the \nicer data to make recommendations for future missions. STROBE-X will observe in the 0.2-12 keV band, using the X-ray Concentrator Array (XRCA), with a planned collecting area of $21760$ cm$^2$ at $1.5$ keV \citep{STROBE_2019}. This increase in collecting area alone will provide a marked improvement over NICER's observations. By conducting tests with longer time spans and smaller uncertainties, we aim to provide thresholds for these future experiments in X-ray timing.

\begin{acknowledgments}
JSH wishes to thank Jim Cordes for useful discussions.  The NANOGrav project receives support from National Science Foundation (NSF) as a Physics Frontier Center award number 1430284.  Portions of this work performed at NRL were supported by NASA.
\end{acknowledgments}



\bibliographystyle{aasjournal}
\bibliography{NicerBib}

\label{lastpage}
\end{document}